\documentclass[aps,prb,twocolumn,groupedaddress,aps]{revtex4}

\usepackage{amsmath}
\usepackage{epsfig}

\begin{document}


\title{Metric-tensor flexible-cell algorithm for isothermal-isobaric
molecular dynamics simulations}

\author{E. Hern\'{a}ndez}
\email[]{ehe@icmab.es}
\affiliation{Institut de Ci\`{e}ncia de Materials 
de Barcelona, ICMAB - CSIC, \\
Campus de Bellaterra, 08193 Bellaterra, Barcelona,
Spain}

\date{\today}

\begin{abstract}
An extended Hamiltonian approach to conduct isothermal-isobaric molecular
dynamics simulations with full cell flexibility is presented. The components
of the metric tensor are used as the fictitious degrees of freedom for the 
cell, thus avoiding the problem of spurious cell rotations and artificial
symmetry breaking effects present in the original Parrinello-Rahman
scheme. This is complemented by the Nos\'{e}-Poincar\'{e} approach for
isothermal sampling. The combination of these two approaches leads to 
equations of motion that are Hamiltonian in structure, and which can 
therefore be solved numerically using recently developed powerful
symplectic integrators. One such integrator, the generalised 
leap-frog, is employed to provide a numerical algorithm for integrating
the isothermal-isobaric equations of motion obtained. 
\end{abstract}

\pacs{02.70.Ns, 31.15.Qg, 71.15.Pd, 83.10.Rs}

\maketitle


\section{Introduction}
\label{sec:introduction}

Over the last few decades advances in techniques, models and computer power
have made simulation methods~\cite{allen:tildesley,frenkel:smit,thijssen}
an indispensable aid to research in condensed
matter, molecular physics and chemistry, and in materials science. By 
emulating the conditions under which experiments are carried out and the 
interactions between the components of the system as closely as possible, 
simulations can often provide information that is not directly attainable 
from the experiments themselves, and can help to interpret the empirical
observations.

There are two large groups of simulation methods that are capable of taking 
thermal effects into account, namely the Monte Carlo~(MC) and the Molecular
Dynamics~(MD) methods. In this work it is the latter that will be 
of interest. Conventional MD consists of numerically integrating the classical 
equations of motion for an ensemble of atoms representative of the system
of interest. Assuming ergodicity, a sufficiently long trajectory will sample
the whole of the accessible phase space under the conditions of 
the simulation, and a time average over such a trajectory of any property
of the simulated system will provide an estimate of the value of that 
property in the real system under the same conditions. The classical 
equations of motion conserve the total energy, so this procedure simulates
the system of interest in the {\em microcanonical\/} 
(constant number of particles, N, constant volume, V, and constant energy, E, 
or NVE) ensemble.
However, experiments are most frequently conducted under conditions of 
constant temperature, and sometimes also constant pressure, conditions which 
correspond to the {\em canonical\/} (NVT) or {\em isothermal-isobaric\/} (NPT)
ensembles respectively. It would therefore be desirable to have MD methods
that were capable of sampling these ensembles also. In a seminal paper,
Andersen~\cite{andersen} proposed two different methods for conducting
MD simulations in the canonical and iso-shape isobaric-isoenthalpic (NPH, where
H is the enthalpy)
ensembles, methods which could be combined to perform simulations sampling
the isothermal-isobaric ensemble. Andersen's NVT~MD method involved a 
series of stochastic {\em collisions\/} which changed the velocity of 
a randomly chosen atom in the system to one generated from a Maxwell-Boltzmann
distribution of velocities at the desired temperature. Such a series of 
collisions changes the total energy of the system so that it fluctuates 
around its equilibrium value as prescribed by the canonical ensemble.
As for the constant-pressure ({\em isobaric-isoenthalpic\/} or NPH) ensemble, 
Andersen showed that it could be sampled by incorporating the volume as 
a new degree of freedom in the classical equations of motion, with a fictitious
mass and velocity associated to it.

Andersen's idea of extending the dynamics of a physical system by including
fictitious degrees of freedom has proved to be extremely powerful. Andersen
himself\cite{andersen} postulated that using this method it 
might be possible to perform
NVT dynamics in a non-stochastic fashion. Indeed, such a method was 
put forward by Nos\'{e}~\cite{nose1,nose2}, and later modified by
Hoover~\cite{hoover}. The same idea was to be used by Parrinello and 
Rahman~\cite{parrinello:rahman,parrinello:rahman_jap} 
for flexible-cell (as opposed to iso-shape)
constant-pressure MD, and by Car and Parrinello~\cite{car:parrinello}, who 
combined in an ingenious way the classical dynamics of ions with the fictitious 
dynamics of the electronic orbitals expanded in a plane-wave
basis set, within a Density Functional Theory description of the electronic
structure. This was done in such a way that as the ions moved, the electronic
orbitals followed adiabatically, so that the system remained in the
ground state, or sufficiently close to it, allowing Car and Parrinello
to carry out {\em ab initio\/}~MD for the first time. 

Although the extended dynamics approach has proved very useful for 
sampling ensembles beyond the microcanonical one, over time a number of 
shortcommings have been detected, both with the Nos\'{e}-Hoover method
for NVT sampling and the Parrinello-Rahman method for NPH simulations.
The method originally proposed by Nos\'{e} involved a time transformation,
as a result of which the phase space is not sampled at regular time 
intervals. While this in itself does not pose a problem for the calculation
of thermal averages of time-independent quantities, it severely complicates
the calculation of time-correlation functions. It is possible to transform
the equations of motion back to real time~\cite{nose2,hoover} using a 
non-canonical~\cite{canonical} transformation, but then the Hamiltonian
structure is lost, i.e. the equations that result cannot be derived from
a Hamiltonian. Although the lack of Hamiltonian structure in the 
Nos\'{e}-Hoover equations poses no practical difficulties (both
iterative~\cite{frenkel:smit} and explicit~\cite{tuckerman:berne:martyna}
integrators can be derived for these equations), it is undesirable from
a formal point of view, not least because the usual machinery of 
statistical mechanics is not directly applicable to non-Hamiltonian
systems (see the recent work of Tuckerman and 
coworkers~\cite{tuckerman:mundy:martyna,tuckerman:liu:ciccotti:martyna}
concerning this point). Recently, Bond, Leimkuhler and 
Laird~\cite{bond:leimkuhler:laird} have shown that it is possible to obtain
NVT sampling (subject to the usual assumption of ergodicity) starting
from Nos\'{e}'s original Hamiltonian, but acted upon by a 
Poincar\'{e} transformation. The transformation is canonical, and therefore
preserves the form of the equations of motion; thus, all the recent 
developments in integration schemes for Hamiltonian 
systems~\cite{sanz-serna:calvo} (see below) can be used to solve 
numerically the resulting equations of motion. Furthermore, the 
Poincar\'{e} transformation is chosen in such a way that the 
sampling of the phase space takes place at regular time intervals. Thus,
the scheme of Bond {\em et al.\/}~\cite{bond:leimkuhler:laird} overcomes
the negative aspects associated with the Nos\'{e}-Hoover method.

Concerning the constant-pressure schemes, 
Parrinello and Rahman formulated their method taking as
extended variables the Cartesian components of the simulation cell vectors,
and constructing a fictitious kinetic energy term from their squared 
velocities. As noted by Cleveland~\cite{cleveland} and by 
Wentzcovitch~\cite{wentzcovitch}, this choice leads to equations of motion
that are not invariant under transformations between equivalent 
cells (modular transformations), leading to unphysical symmetry breaking
effects. An added difficulty is that the fictitious dynamics based on 
this choice of variables results in spurious cell rotations, which 
complicate the analysis of the results. Numerous proposals to overcome
these difficulties have appeared in the 
literature~\cite{cleveland,wentzcovitch,lill,melchionna,martyna,souza:martins},
but here I am going to focus on the method proposed by Souza and
Martins~\cite{souza:martins}, which uses as fictitious dynamical
variables the components of the cell metric tensor. The metric tensor
is independent of the cell orientation, and by phrasing the dynamics in 
terms of it, spurious cell rotations simply do not appear. Furthermore, 
the fictitious kinetic energy associated with the metric tensor can
easily be constructed such that it is invariant with respect to modular
transformations.

Important advances have also been achieved in the understanding and design
of integrators for dynamical 
systems~\cite{sanz-serna:calvo,tuckerman:berne:martyna}. Classical
Hamiltonian dynamics is time reversible (if the Hamiltonian is an even
function of the momenta) and symplectic, i.e. it preserves the sum of areas
spanned by the vector products $dp_i \times dq_i$, the area element around
the point $(p_i, q_i)$. It is desirable that numerical integrators
for Hamiltonian equations of motion respect the symmetries underlying
Hamiltonian dynamics, such as time reversibility and symplecticness, 
as then one can be more confident that the discrete-time solution will
resemble more closely the exact solution. Thus considerable efforts
have been devoted to the development of integrators which comply with these 
requirements, and modern simulation techniques should take advantage of
the progress achieved in this field.

Constant-pressure algorithms such as the method of Parrinello-Rahman 
and its variants sample the isobaric-isoenthalpic ensemble. However,
this ensemble is not very common, nor indeed is it very convenient.
In many circumstances it is desirable to have control over the average 
temperature, and it is therefore preferable to perform a simulation
which samples the isothermal-isobaric ensemble. This can be achieved by
combining one of the constant-pressure algorithms with a Nos\'{e}-type
thermostat.
The purpose of this paper is to describe how the metric-tensor based
scheme of Souza and Martins~\cite{souza:martins} for constant-pressure
simulations can be combined with the Nos\'{e}-Poincar\'{e} scheme
of Bond {\em et al.\/}~\cite{bond:leimkuhler:laird} to provide an extended
Hamiltonian for isothermal-isobaric MD simulations with full-cell
flexibility, incorporating the advantages of both methods over the 
original schemes of Nos\'{e}-Hoover and Parrinello-Rahman. 
Similar schemes have been recently presented by 
Sergi {\em et al.\/}~\cite{sergi} and by Sturgeon and 
Laird~\cite{sturgeon:laird}, but only in the case of isotropic cell
fluctuations. It is shown how the equations of motion that result in 
the present scheme can be integrated numerically 
using the {\em generalised leap-frog\/} method~\cite{hairer,sun}, leading
to a symplectic, highly stable algorithm. This NPT MD algorithm is then 
illustrated with a series of realistic test cases, namely NPT simulations
of diamond and crystalline silicon. 

The structure of the paper is as 
follows: in section~\ref{sec:methodology} the necessary background on
the metric-tensor constant-pressure and Nos\'{e}-Poincar\'{e} isothermal
formalisms is reviewed, both schemes are then combined, and a recipe
for integrating the resulting equations of motion numerically is provided.
The methodology is then applied to the test cases in 
section~\ref{sec:applications}, and the conclusions are summarized
in section~\ref{sec:conclusions}.

\section{Methodology}
\label{sec:methodology}

Classical Lagrangian and Hamiltonian mechanics\cite{goldstein}
provide elegant theoretical frameworks in which to set up equations of 
motion for the system under study, using the variables or coordinates
most convenient in each situation. The {\em generalized leap-frog\/} 
integration scheme that will be used below is most readily applied
to Hamiltonian equations, and therefore the Hamiltonian formulation of classical
mechanics will be used in what follows, although the Lagrangian treatment
is of course totally equivalent.

When considering a system in which both the volume and shape of the simulation
cell are allowed to evolve in time, it is helpful to take into account the 
covariant/contravariant character of the dynamical variables.
Therefore, in the following, the covariant or contravariant character 
of the different variables that will be used shall be
indicated by means of a subindex or superindex, respectively.

Let us now review the set of dynamical variables which will be used to 
specify the state of the system. Firstly, it will be most convenient to 
use {\em lattice coordinates\/}, ${\bf q}^i$, which give the position
of atom {\em i\/} relative to the simulation cell. The dynamics of atoms is
then fully accounted for by considering the momenta ${\bf p}_i$, conjugate
to ${\bf q}^i$. Lattice coordinates are related to the usual Cartesian
ones, ${\bf r}_i$, through
\begin{eqnarray}
{\bf r}_i = {\bf H} {\bf q}^i,
\label{eq:lattice_to_cartesian}
\end{eqnarray}
where ${\bf H}$ is a $3\times 3$ matrix formed by the simulation cell vectors
${\bf a}_\alpha, (\alpha = 1,2,3)$ in columns. Since the cell vectors 
${\bf a}_\alpha$ are linearly independent, matrix ${\bf H}$ can be inverted,
and its inverse, ${\bf H}^{-1}$, has row $\alpha$ equal to ${\bf b}^\alpha$,
the reciprocal vector of ${\bf a}_\alpha$ in the following sense:
\begin{eqnarray}
{\bf a}_\alpha \cdot {\bf b}^\beta = \delta_\alpha^\beta.
\end{eqnarray}
The calculation of interatomic distances when atom positions are specified
in terms of lattice coordinates is given by:
\begin{eqnarray}
r_{ij} = \sqrt{({\bf q}^i - {\bf q}^j) \, {\bf G} \, ({\bf q}^i - {\bf q}^j)},
\label{eq:distance}
\end{eqnarray}
where ${\bf G}$ is the metric tensor, with elements
\begin{eqnarray}
G_{\alpha\beta} = {\bf a}_\alpha \cdot {\bf a}_\beta.
\end{eqnarray}
The volume of the simulation cell is also given by the metric tensor as
$V_{cell} = \sqrt{\det {\bf G}}$. The set of vectors 
${\bf b}^\alpha$ also define a metric tensor, $G^{\alpha\beta}$, which is 
reciprocal to the previous one.

Souza and Martins\cite{souza:martins} have shown that the metric tensor
constitutes a very convenient dynamical variable for constant-pressure
MD simulations. Firstly, ${\bf G}$ is invariant under cell rotations; the
orientation of the cell is irrelevant, and thus spurious cell rotations do
not appear during the dynamics. Secondly, it is easy to set up a fictitious
kinetic energy term associated with the metric tensor (see below) which 
is invariant under modular transformations\cite{wentzcovitch} (i.e. 
transformations between the different possible cells compatible with the
periodicity of the system). This avoids artificial symmetry breaking effects.
Thus, spurious cell rotations and symmetry breaking effects, which 
appeared in the original 
Parrinello-Rahman\cite{parrinello:rahman,parrinello:rahman_jap}
constant-pressure algorithm, are naturally avoided in this formalism.
Following Souza and Martins, each metric tensor component 
$G_{\alpha\beta}$ has a conjugate momentum $P^{\alpha\beta}$, and the 
fictitious kinetic energy term associated to the dynamics of the metric
tensor is
\begin{eqnarray}
K_G = \frac{P^\alpha_{\ \beta} \, P^\beta_{\ \alpha}}{2 M_G \det{\bf G}},
\label{eq:metric_K}
\end{eqnarray}
where the sum over repeated indices is implied. Here $M_G$ is a fictitious
mass, but the total effective mass is $M_G \det{\bf G}$, which varies
with the cell volume. While it would be possible to use a constant 
fictitious mass, this form has the particularity of reducing the kinetic
energy expression in Eq.~(\ref{eq:metric_K}) to the same form as
in Andersen's constant-pressure scheme in the case of iso-shape cell 
fluctuations.

In the case of hydrostatic external pressure, the potential energy term
associated to the metric dynamics is simply
\begin{eqnarray}
{\cal U}_G = {\cal P}_{ext} \, V_{cell} = 
      {\cal P}_{ext} \sqrt{\det {\bf G}}.
\end{eqnarray}
Souza and Martins went on to show that the case of an anisotropic external
stress can also be contemplated, if a potential energy term of the form
\begin{eqnarray}
{\cal U}_G = \frac{1}{2} \sigma_{ext}^{\beta\alpha}\, G_{\alpha\beta}
\end{eqnarray}
is included, where $\sigma_{ext}^{\beta\alpha}$ are the components
of the external stress in contravariant lattice coordinates.

The combined dynamics of atoms and metric tensor described so far conserves
the enthalpy (the generalized enthalpy of Thurston\cite{thurston}, in the
case of constant anisotropic external stress), and samples the 
isobaric-isoenthalpic (NPH) ensemble. 
It is therefore desirable to combine
the dynamics of the extended system of atoms and metric tensor with a 
device that enables sampling of the isobaric-isothermal (NPT) ensemble.
Souza and Martins used stochastic Langevin dynamics in the examples 
reported in their work. A different strategy will be pursued here, which
consists of coupling the dynamics of atoms and metric tensor with a 
Nos\'{e} thermostat, as described in what follows.

Canonical (NVT) MD simulations have been usually undertaken by means
of the so called Nos\'{e}-Hoover method. Recently, however, 
Bond {\em et al.\/}~\cite{bond:leimkuhler:laird} have provided an alternative
scheme, which also samples the NVT ensemble, but has the additional 
advantage of being Hamiltonian in structure, thereby permitting the use
of symplectic numerical integrators. This is achieved by performing a 
Poincar\'{e} transformation on the original Nos\'{e} Hamiltonian,
$H_{\mbox{\scriptsize Nos\'{e}}}$, which results in
\begin{eqnarray}
H_{\mbox{\scriptsize Nos\'{e}-Poincar\'{e}}} = 
        S ( H_{\mbox{\scriptsize Nos\'{e}}} - H_0),
\label{eq:nose-poincare}
\end{eqnarray}
where $H_0$ is a suitably chosen constant, 
and $H_{\mbox{\scriptsize Nos\'{e}}}$ is given by
(in Cartesian coordinates)
\begin{eqnarray}
H_{\mbox{\scriptsize Nos\'{e}}} = \sum_i \frac{p^2_i}{2\, m_i\, S^2} + 
       {\cal U}({\bf r})
	+ \frac{P_S^2}{2 M_S} + g \, k_B \, T_{ext} \ln S.
\end{eqnarray}
Here $S$ is the position variable of the thermostat, a strictly
positive quantity, $P_S$ its 
conjugate momentum, $g$ is the number of degrees of freedom of the 
physical system (i.e. excluding extended or fictitious dynamical variables),
$k_B$ is Boltzmann's constant, and $T_{\mbox{\em ext}}$ is the temperature
of the thermostat. Bond and coworkers have demonstrated that, under
the assumption of ergodicity, the
Nos\'{e}-Poincar\'{e} Hamiltonian generates dynamics that sample the 
canonical ensemble, as desired. Recently, Sturgeon and
Laird\cite{sturgeon:laird} have extended the Nos\'{e}-Poincar\'{e} Hamiltonian
with an Andersen barostat, which implements iso-shape cell fluctuations,
thereby sampling the isothermal-isobaric (NPT) ensemble appropriate
for non-crystalline systems. For crystalline solids, however, it is necessary
to allow fluctuations of the cell shape as well as of its volume, which 
can be done by means of the metric tensor dynamics described above. The
scheme presented here can therefore be regarded as a generalization of
the method reported by Sturgeon and Laird to full-cell dynamics
NPT MD simulations.

By combining the Nos\'{e}-Poincar\'{e} Hamiltonian 
[Eq.~(\ref{eq:nose-poincare})] of Bond and 
coworkers\cite{bond:leimkuhler:laird} with the metric tensor constant-pressure
scheme of Souza and Martins\cite{souza:martins}, one arrives at the 
following Hamiltonian:
\begin{eqnarray}
H_{\rm\scriptsize NPT} = S \left[ \sum_i 
 \frac{p_{i\alpha} \, p_i^{\alpha}}{2 \, m_i\, S^2} + 
     {\cal U}({\bf q},{\bf G}) + 
    \frac{P^{\alpha}_{\ \beta}\, P^\beta_{\ \alpha}}{2 M_G \det {\bf G}} \: + 
     \right. \nonumber \\ {\cal P}_{ext}\, \sqrt{\det {\bf G}} \: + 
     \frac{1}{2} \sigma_{ext}^{\beta\alpha} \, G_{\alpha\beta} \: + \\ \left.
     \frac{P_S^2}{2 M_S} + g\, k_B T_{ext}\ln S - H_0 \right] \nonumber .
\label{eq:NPTHamiltonian}
\end{eqnarray}
In this equation, lower case labels {\em p\/} and {\em q\/} refer to atomic
momenta and position variables of the physical system of interest, while labels 
in upper case refer to fictitious degrees of freedom (thermostat or 
barostat). Latin indices label atoms, always appear first and are always
used as subindices, while Greek indices label components of tensors, and if used
as subindices appear after the atom label. 
Sums over atoms, as in the atomic kinetic energy term, are written out
explicitly; otherwise the summation over repeated indices of tensorial 
quantities is implied.
Thus, $p_{i\alpha}$ is the
covariant $\alpha$ component of the momentum of atom {\em i\/}, while
$P^\alpha_{\ \beta}$ is the mixed (contravariant-covariant) component of the 
second-rank tensor $P_{\alpha\beta}$ formed by the momenta associated to 
the components of the metric tensor {\bf G}.  The constant $H_0$ is to be 
chosen so that $H_{\rm NPT}$ has a value of zero. 

Using the standard rules of Classical Mechanics\cite{goldstein}, and with
the help of the relations $\partial \det {\bf G}/\partial G_{\alpha\beta} = 
G^{\beta\alpha} \det {\bf G}$ and $\partial G^{\lambda\mu}/ \partial 
G_{\alpha\beta} = -G^{\lambda\alpha} G^{\beta\mu}$, it is straight forward
to obtain the following equations of motion:
\begin{widetext}
\begin{subequations}
\begin{eqnarray}
\dot{q}_i^\alpha & = & \frac{p_i^\alpha}{m_i\: S} \\
\dot{p}_{i\alpha}  & = & -S \: \frac{\partial {\cal U}}{\partial q_i^\alpha} \\
\dot{G}_{\alpha\beta} & = &  S \: \frac{P_{\alpha\beta}}{M_G \: \det {\bf G}} \\
\dot{P}^{\alpha\beta} & = & -S\left[ 
         \frac{\partial {\cal U}}{\partial G_{\alpha\beta}}
        - \sum_i \frac{p_i^\alpha \: p_i^\beta}{2\, m_i\: S^2} +
  \frac{P^{\beta\lambda} \, G_{\lambda\mu}\, P^{\mu\alpha}}{M_G \det {\bf G}} 
        \right. \nonumber \\ 
        & &\hspace{3.0cm} \left.
       + \left(\frac{1}{2} {\cal P}_{ext}\, \sqrt{\det {\bf G}} \: - 
 \frac{P^\lambda_{\ \mu} \: P^\mu_{\ \lambda}}{2\, M_G \: \det {\bf G}} \right)
    G^{\beta\alpha}  + \frac{1}{2} \sigma_{ext}^{\beta\alpha}\right]  \\
\dot{S} & = & S \, \frac{P_S}{M_S} \\
\dot{P}_S & = & \sum_i \frac{p_{i\alpha}\: p_i^\alpha}{m_i\, S^2} - 
    g\, k_B T_{ext} - \Delta H 
\label{eq:NPTequations}
\end{eqnarray}
\end{subequations}
\end{widetext}
where the dot indicates a time derivative, and
\begin{eqnarray}
\Delta H = \sum_i \frac{p_{i\alpha} \: p_i^\alpha}{2\, m_i\: S^2} + 
   {\cal U}({\bf q},{\bf G}) + 
      \frac{P^\lambda_{\ \mu} \: P^\mu_{\ \lambda}}{2\, M_G \:
   \det {\bf G}}  \nonumber \\
   + {\cal P}_{ext} \, \sqrt{\det {\bf G}} + 
    \frac{1}{2} \sigma_{ext}^{\beta\alpha} G_{\alpha\beta} + \\
       g\, k_B T_{ext} \ln S - H_0. \nonumber
\end{eqnarray} 

Let us now turn to the question of how to obtain a numerical scheme to
integrate these equations of motion. Of the different schemes that are 
possible, I have chosen to
use the {\em generalized leap-frog scheme\/}~(GLF)\cite{hairer,sun}, which
is simple, and since the system is Hamiltonian\cite{sanz-serna:calvo},
the numerical procedure that results from applying the GLF to 
Eqs.~(\ref{eq:NPTequations}) is symplectic and 
time-reversible\cite{bond:leimkuhler:laird,sturgeon:laird}. Let us briefly
recall how the GLF works: let $Q, P$ represent position and momentum
variables respectively. The Eqs.~(\ref{eq:NPTequations}) can be 
generally put in the form
\begin{eqnarray}
\dot{Q} = G(P,Q), \\
\dot{P} = F(P,Q).
\end{eqnarray}
The GLF scheme consists of first propagating the momentum variables half a 
time-step forward from the initial point in phase space:
\begin{eqnarray}
P(t&+&1/2\:\delta t) = P(t) + \nonumber \\ & \delta t & \,
       F[P(t+1/2\:\delta t), Q(t)]/2,
\label{eq:first-step}
\end{eqnarray}
where $\delta t$ is the time step.
Note that the new momenta appear on both sides of the equation, which can lead
(and in general, does lead) to implicit equations for the updated
momentum variables. Next, the position variables are updated a full time-step:
\begin{eqnarray}
Q(t + \delta t) &=& Q(t) + \nonumber \\ & \delta t & \left\{  
       G[P(t+1/2\: \delta t), Q(t)] + \right. \\ & &
       \left. G[P(t+1/2\: \delta t), Q(t + \delta t)]\right\}/2. \nonumber
\end{eqnarray}
Again, due to the fact that $Q(t + \delta t)$ appears on both sides,
in the general case one obtains an implicit equation, like for the momentum
variables at half step.  Once $Q(t + \delta t)$ has been obtained, 
new forces can be calculated, and thus it is possible to
bring up the momenta to full step:
\begin{eqnarray}
P(t&+&\delta t) = P(t+1/2\:\delta t) + \nonumber \\ & \delta t & \,
       F[P(t+1/2\: \delta t), Q(t+\delta t)]/2.
\end{eqnarray}
The scheme can then be iterated by setting $t = t + \delta t$ and returning to
Eq.~(\ref{eq:first-step}).

Applying the GLF to the NPT equations of motion~(\ref{eq:NPTequations})
leads to the following numerical scheme:
\begin{widetext}
\begin{subequations}
\begin{eqnarray}
p_{i\alpha,1/2} & = & p_{i\alpha,0} - 
 \frac{1}{2} \delta t\, S_0 
   \frac{\partial {\cal U}}{\partial q_{i,0}^{\alpha}}  \\
P^{\alpha\beta}_{1/2} & = & P^{\alpha\beta}_0 - 
  \frac{1}{2} \delta t \: S_0 \left[\frac{\partial {\cal U}}{\partial 
     G_{\alpha\beta,0}} - 
\sum_i\frac{p_{i,1/2}^\alpha \: p_{i,1/2}^\beta}
{2\, m_i \: S^2_0} + \frac{P^{\beta\lambda}_{1/2}\:
  G_{\lambda\mu,0} \: P^{\mu\alpha}_{1/2}}{M_G\: \det{\bf G}_0}
  \right. \nonumber \\ 
  & & \hspace{3.0cm} \left.
  + \left(\frac{1}{2} {\cal P}_{ext} \: \sqrt{\det {\bf G}_0} - 
\frac{P^\lambda_{\ \mu,1/2} \: P_{\ \lambda,1/2}^\mu}
{2\, M_G\: \det{\bf G}_0}\right) G^{\beta\alpha}_0 + 
     \frac{1}{2} \sigma_{ext}^{\beta\alpha} \right]  \\
P_{S,1/2} & = & P_{S,0} + \frac{1}{2}\delta t \left[ 
  \sum_i\frac{p_{i\alpha,1/2} \: p_{i,1/2}^\alpha}{m_i\, S^2_0}
  - g\, k_B T_{ext} \: - \right. \nonumber \\
  & & \hspace{3.0cm} \left.  \Delta H({\bf q}_0, {\bf G}_0, S_0, {\bf p}_{1/2},
     {\bf P}_{G,1/2}, P_{S,1/2})  \right] \\ 
S_1 & = & S_0 + \frac{1}{2} \delta t \left( S_0 \frac{P_{S,1/2}}{M_S} +
           S_1 \frac{P_{S,1/2}}{M_S}\right) \\ 
G_{\alpha\beta,1} & = & G_{\alpha\beta,0} + \frac{1}{2} \delta t
   \left( S_0 \frac{G_{\beta\lambda,0} \: P^{\lambda\mu}_{1/2} \: 
       G_{\mu\alpha,0}}{M_G \: \det {\bf G}_0} + 
 S_1 \frac{G_{\beta\lambda,1} \: P^{\lambda\mu}_{1/2} \:
              G_{\mu\alpha,1}}{M_G \: \det {\bf G}_1} \right) \\
q_{i,1}^\alpha & = & q_{i,0}^\alpha + \frac{1}{2}\delta t 
      \left(\frac{p_{i,1/2}^\alpha}{m_i \: S_0} +
            \frac{p_{i,1/2}^\alpha}{m_i \: S_1}\right) \\
P_{S,1} & = & P_{S,1/2} + \frac{1}{2} \delta t \left[
  \sum_i \frac{p_{i\alpha,1/2} \: p_{i,1/2}^\alpha}{m_i \:  
        S^2_1} - g\, k_B T_{ext} \: - \right. \nonumber \\ 
        & & \hspace{3.0cm} \left. 
        \Delta H({\bf q}_1, {\bf G}_1, S_1,{\bf p}_{1/2},
        {\bf P}_{G,1/2}, P_{S,1/2}) \right] \\ 
P^{\alpha\beta}_1 & = & P^{\alpha\beta}_{1/2} - \frac{1}{2} \delta t \:
 S_1 \left[ \frac{\partial {\cal U}}{\partial G_{\alpha\beta,1}}
 -\sum_i\frac{p_{i,1/2}^\alpha \: p_{i,1/2}^\beta}{2 \, m_i \: S^2_1}
 +\frac{P^{\beta\lambda}_{1/2} \: G_{\lambda\mu,1} \:
  P^{\mu\alpha}_{1/2}}{M_G\: \det{\bf G}_1} \right.  \nonumber \\ 
         & & \hspace{3.0cm} \left.
  + \left(\frac{1}{2}{\cal P}_{ext} \, \sqrt{\det {\bf G}_1} -
  \frac{P^\lambda_{\ \mu,1/2} \: P_{\ \lambda,1/2}^\mu}
  {2 \, M_G \: \det{\bf G}_1}\right) G^{\beta\alpha}_1 + 
   \frac{1}{2} \sigma_{ext}^{\beta\alpha}\right] \\
p_{i\alpha,1} & = & p_{i\alpha,1/2} - \frac{1}{2} \delta t \:
     S_1 \frac{\partial {\cal U}}{\partial q_{i,1}^\alpha}
\label{eq:numericalNPT}
\end{eqnarray}
\end{subequations}
\end{widetext}

In these equations, subindices of 0, 1/2 and 1 indicate that the corresponding
dynamical variable should be taken at zero, half or full time step from the
initial point in phase space. 
Eqs.~(\ref{eq:numericalNPT}a-c) propagate the momenta forward in time
by half a time step; Eqs.~(\ref{eq:numericalNPT}d-f) advance the position
variables the whole length of the time step, and finally 
Eqs.~(\ref{eq:numericalNPT}g-i) complete the updating of the momenta
to full time-step. The order in which Eqs.~(\ref{eq:numericalNPT}) are
written corresponds to the order in which they must be implemented
in a computer code.
Note that, as indicated above, some of the
equations are implicit. In particular, Eq.~(\ref{eq:numericalNPT}b) 
is implicit in the metric tensor momenta at half time step, and must be
solved iteratively. This is also the case for the metric tensor components
themselves, in Eq.~(\ref{eq:numericalNPT}e). Two strategies can be
adopted for solving these equations: a Newton-Raphson\cite{numerical_recipes}
procedure can be easily applied, but a simple iterative scheme seems to
work equally well and is even simpler to program. 
In the example applications reported below, the iterative scheme was used.
This consisted of using as initial guess for the half-step metric-tensor
momenta in Eq.~(\ref{eq:numericalNPT}b) ($P^{\alpha\beta}_{1/2}$) their
values at the start of the molecular dynamics time step
(i.e.~$P^{\alpha\beta}_{0}$). The resulting values for 
$P^{\alpha\beta}_{1/2}$ are then fed again into Eq.~(\ref{eq:numericalNPT}b),
thus obtaining a new estimate for the correct values at half-step. This
procedure is iterated until the absolute values of the differences
of $P^{\alpha\beta}_{1/2}$ found in two successive iterations of the procedure 
differ by less than $10^{-7}$. Since, for a sufficiently small time step,
the values of $P^{\alpha\beta}_{0}$ are close to those of 
$P^{\alpha\beta}_{1/2}$, this procedure converges very rapidly, requiring
only a few iterations. Exactly the same procedure was followed for
solving Eq.~(\ref{eq:numericalNPT}e).
It should be emphasized
that the implicit nature of Eqs.~(\ref{eq:numericalNPT}b)
and~(\ref{eq:numericalNPT}e) does not have any significant impact on 
performance; the iterative procedure is very fast, and does not require
re-evaluation of the energy, force or stress; however, because of the fact
that a convergence criterion must be used, the exact time-reversibility of
the equations of motion [Eqs.~(\ref{eq:NPTequations})] is lost in the
numerical scheme of Eqs.~(\ref{eq:numericalNPT}). Nevertheless, the use of
a strict convergence criterion makes the scheme time-reversible within
numerical accuracy.
Equation~(\ref{eq:numericalNPT}c) also deserves attention; it is quadratic
in $P_{S,1/2}$, and care must be taken to choose the right root and avoid
numerical cancellation errors\cite{bond:leimkuhler:laird}.

Equations~(\ref{eq:numericalNPT}) are quite simple to incorporate into 
existing MD codes. The only additional information required
from what is conventionally computed in an MD program are the derivatives
of the potential energy with respect to the metric tensor components,
$\frac{\partial {\cal U}}{\partial G_{\alpha\beta}}$. These can be 
obtained taking into account that inter atomic distances depend on the
components of the metric tensor, as indicated in Eq.~(\ref{eq:distance}).
However, if the program already computes the derivatives 
$\frac{\partial {\cal U}}{\partial \varepsilon_{\lambda\mu}}$, where
$\varepsilon_{\lambda\mu}$ are the components of the Cartesian strain tensor,
this information can be converted into the required derivatives in
a straight forward fashion, as the strain and metric tensors 
are related through\cite{parrinello:rahman_jap}
\begin{eqnarray}
{\bf \varepsilon} = \frac{1}{2} \left[ ({\bf H}^{-1}_0)^T {\bf G} \,
       {\bf H}^{-1}_0 - {\bf 1} \right],
\end{eqnarray}
where $\bf H_0^{-1}$ is the inverse of the undistorted cell matrix, and {\bf 1} 
is the unit matrix. Then, by simple application of the chain rule, one obtains
the following relation:
\begin{eqnarray}
\frac{\partial {\cal U}}{\partial G_{\alpha\beta}} = \frac{1}{2} 
(H_0^{-1})_{\alpha\lambda} 
\frac{\partial {\cal U}}{\partial \varepsilon_{\lambda\mu}} 
(H_0^{-1})_{\mu\beta}^T,
\end{eqnarray}
where again summation over repeated indices is implied.

\section{Applications}
\label{sec:applications}

The metric tensor Nos\'{e}-Poincar\'{e} NPT method described in
section~\ref{sec:methodology} has been implemented in a computer 
program which performs MD simulations of systems described with a
Tight Binding\cite{tb_review}~(TB) total energy model. In this section,
after a brief description of the model used, I will report the results
of tests of the methodology and some examples of applications directed
at demonstrating its usefulness.

\subsection{Model}
\label{sub:model}

The NPT algorithm described in this paper applies equally well to any atomistic
model from which forces and stresses, as well as the total energy, can
be obtained, and therefore any such models can be used in conjunction with
it. Atomistic models can be broadly classified into three groups,
namely {\em empirical potential\/} methods, {\em semi-empirical\/}
total energy methods and {\em first principles\/} or {\em ab initio\/} total
energy methods. Each group of methods has its strengths and weaknesses;
for example, empirical potential methods are computationally cheap, but
{\em ad hoc\/} in nature, and therefore of limited reliability and
transferability.  First principles
methods, on the other hand, rest on firm theoretical grounds, but are
orders of magnitude more demanding computationally than empirical 
potentials. Added difficulties here are the bad scaling of the 
computational cost with the size of the system under study (usually, for
electronic structure methods
the cost scales as $N^3$ or worse, where $N$ is the number of atoms, although
considerable progress towards linear-scaling methods\cite{goedecker,ordejon}
has been achieved in recent years), and, of specific relevance 
to constant-pressure simulations,
the accuracy of the total energy and its derivatives depends on the volume
of the simulation cell, unless highly converged integrations over the
irreducible cell of the reciprocal lattice are used\cite{payne}.

In the applications that follow, I have chosen to use an approximate
Tight Binding~(TB) model\cite{tb_review}, which corresponds to the group of 
semi-empirical methods discussed above. This choice is motivated by
several considerations. Being in between the extremes of
empirical potentials and first principles methods, it shares some of 
the advantages (but also some of the disadvantages) of both. Its computational
demands are very modest; it is based on a quantum treatment of the 
valence electrons, even if at a rather simplified level, and therefore it 
incorporates the essential features of the quantum nature of the chemical
bond. In spite of their simplicity, there exist TB models which are capable
of surprising accuracy in their predictions. In particular, the applications
reported below have been carried out using a TB model due to
Porezag and coworkers\cite{porezag}. It goes beyond conventional TB models
in that it incorporates the non-orthogonality of the basis set, which is
usually assumed to be orthogonal, is constructed on the basis of Density
Functional calculations employing the same basis set, and the range of
the hopping integrals extends beyond the nearest neighbor distance, which
is the range used in most TB models. Additional details on this model
can be found in the paper by Porezag {\em et al.\/}\cite{porezag}. 

\begin{figure}[t]
\begin{center}
\leavevmode
\epsfxsize=8.5cm
\epsffile{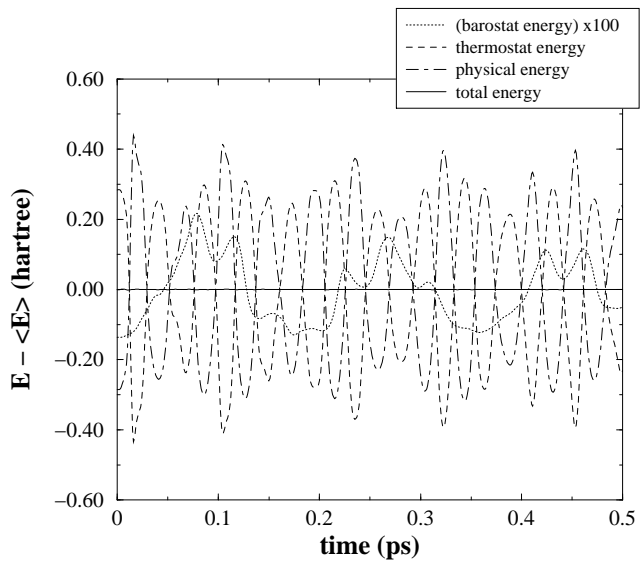}
\end{center}
\caption{Contributions to the conserved quantity $H_{\rm\scriptsize NPT}$
[Eq.~(\ref{eq:NPTHamiltonian})] for a carbon system in the diamond
structure (54-atom cell) as a function of time. The curve labeled
{\em physical energy\/} is the energy of the atomic sub-system
(kinetic plus potential energy).}
\label{fig:energies}
\end{figure}

\subsection{Diamond}
\label{sub:diamond}

To illustrate the stability and accuracy of the integration scheme embodied
in Eqs.~(\ref{eq:numericalNPT}a-i), a simulation of diamond at 
0~GPa external pressure and a temperature of 1000~K
was performed. This is well below the Debye temperature of 
diamond (2340~K), and therefore it is strictly speaking not justified to 
perform a classical MD simulation at this temperature. Nevertheless the
aim here is to test the methodology, and not to extract any conclusions
on the physics of diamond at 1000~K. The starting
configuration of the system consisted of~54 atoms at their equilibrium 
positions in a cell of edge length 14.33~bohr (7.58~\AA), with the edges forming
$60^o$~angles. Initial velocities were chosen randomly from the
Maxwell-Boltzmann distribution at the desired temperature, and modified to
eliminate any translation of the center of mass. The fictitious mass of the
Nos\'{e} thermostat was chosen to be equal to the mass of a carbon atom, while
for the barostat a mass of 10~au was used. Both thermostat and barostat
were assigned zero initial momenta, and the value of $H_{\rm\scriptsize NPT}$
at zero time fixes the value of $H_0$.

The total energy of the system [${\cal U}({\bf q},{\bf G})$ in 
Eq.~(\ref{eq:NPTHamiltonian})] was calculated using the TB~model of 
Porezag and coworkers~\cite{porezag} described above. The generalized 
eigenvalue equation that results for each system configuration according to
this model was solved using direct diagonalisation, and a set of~4
reciprocal lattice vectors chosen according to the Monkhorst-Pack
scheme~\cite{monkhorst:pack} were used, which were sufficient to converge
the total energy better than 0.04~meV/atom. The length of the time step
was set to 1~fs, and the simulation was run for a total of 10~ps.


\begin{figure}[t]
\begin{center}
\leavevmode
\epsfxsize=8.5cm
\epsffile{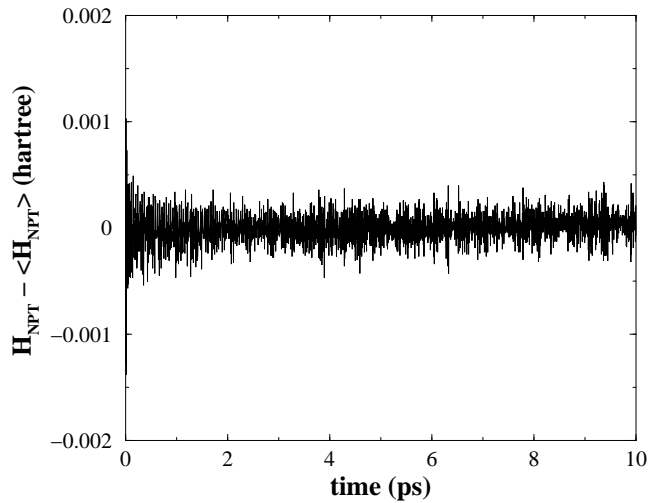}
\end{center}
\caption{$H_{\rm\scriptsize NPT}$ [Eq.~(\ref{eq:NPTHamiltonian})]
as a function of time for the diamond system.}
\label{fig:Hnpt}
\end{figure}

In Fig.~(\ref{fig:energies}) the different terms contributing to 
$H_{\rm\scriptsize NPT}$ [Eq.~(\ref{eq:NPTHamiltonian})] have been
plotted as a function of time for the first 0.5~ps of the simulation. 
The {\em physical energy\/} is the sum of 
the kinetic and potential energy terms, calculated from the 
atomic momenta and from ${\cal U}({\bf q},{\bf G})$,
respectively. The thermostat and barostat contributions include both
the kinetic and potential energy terms associated to each of these 
fictitious degrees of freedom (the barostat energy has been scaled by 
a factor of 100 so that it can be appreciated on the graph). Note that, 
as expected, the total energy of the atoms (their kinetic plus potential
energy, the physical energy), is not conserved, contrary to
what would happen in a conventional microcanonical (NVE)~MD simulation.
In the present case, it is $H_{\rm\scriptsize NPT}$ that is the 
conserved quantity, and as can be seen from Fig.~(\ref{fig:energies}), this
is indeed approximately conserved by the numerical scheme of
Eqs.~(\ref{eq:numericalNPT}a-i). To judge how well 
$H_{\rm\scriptsize NPT}$ is conserved during the dynamics, it has been
plotted in Fig.~(\ref{fig:Hnpt}) for the whole length of the simulation.
Except at the very first stages of the calculation, the deviations of
$H_{\rm\scriptsize NPT}$ from its mean value are smaller than 0.0005~hartree,
and its standard deviation has been computed to be 0.00015~hartree. No
drift is observed during the trajectory; simulations of up to 50~ps
(see below) were also carried out, and again, no appreciable drift was
observed, testifying to the stability of the method. 

\begin{figure}[t]
\begin{center}
\leavevmode
\epsfxsize=8.5cm
\epsffile{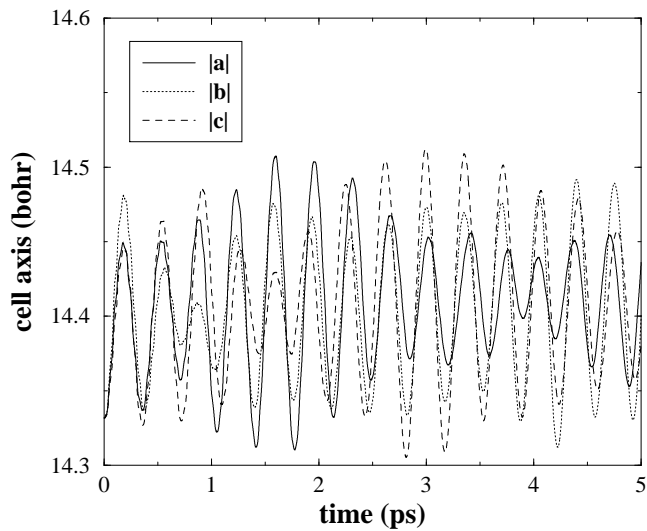}
\end{center}
\caption{Cell vector lengths as a function of time for the 54-atom diamond
cell.}
\label{fig:cell-axis}
\end{figure}

\begin{figure}[t]
\begin{center}
\leavevmode
\epsfxsize=8.5cm
\epsffile{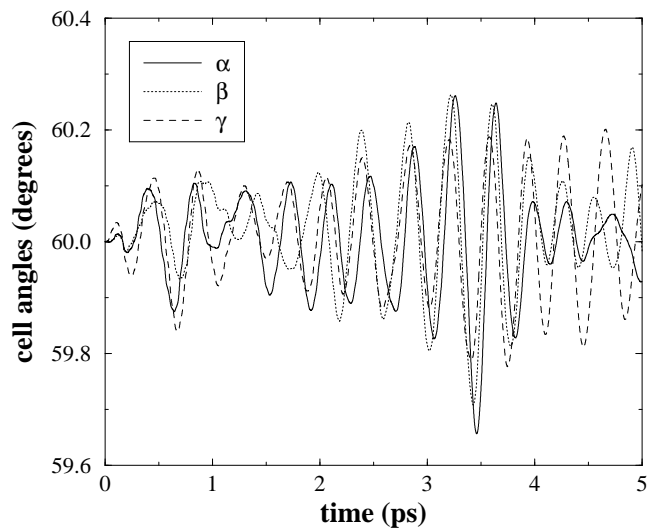}
\end{center}
\caption{Cell angles as a function of time for the 54-atom diamond cell.}
\label{fig:cell-angles}
\end{figure}

Figs.~(\ref{fig:cell-axis}) and~(\ref{fig:cell-angles}) illustrate the
time evolution of the super-cell edge lengths and angles during
the first 5~ps of the simulation, respectively. The edges start at time 
zero having equal lengths, corresponding to that of a 54-atom cell at 0~K,
but as the simulation proceeds each cell parameter evolves separately. Note 
how the average values of the moduli of the cell vectors settle at a 
higher value than the 0~K one, namely at 14.41~bohr (obtained by averaging
over the whole length of the simulation). This is due to the thermal 
expansion of diamond from 0 to 1000~K, though it should be emphasized
that one must not expect an accurate estimation of the thermal 
expansion in this case, partly because of the complete neglect of quantum
effects, which, as indicated above, are important below the Debye temperature. 
The cell angles also evolve independently shortly after the start of the 
simulation, but contrary to what happens with the cell vector moduli, 
which evolve to a different mean value, the cell angles oscillate around
their initial value of~$60^o$, with their instantaneous values remaining
within $\pm 0.4^o$, indicating that, although the cell expands due to the
thermal motion of the atoms at 1000~K, it does not change its shape.

\begin{figure}[b]
\begin{center}
\leavevmode
\epsfxsize=8.5cm
\epsffile{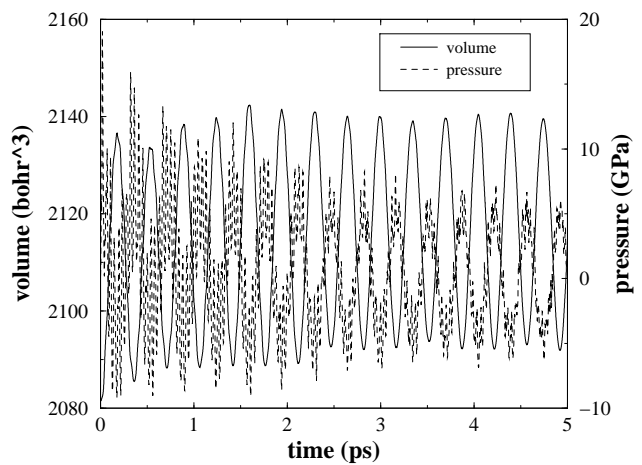}
\end{center}
\caption{Cell volume (left ordinate axis) and internal pressure (right ordinate
axis) for the 54-atom diamond cell.}
\label{fig:volume}
\end{figure}

Fig.~(\ref{fig:volume}) illustrates the time evolution of the cell volume
and the internal pressure during the first 5~ps of the simulation. As expected,
these two magnitudes display opposite behavior, in the sense that when
the volume is lowest, the internal pressure is highest, and vice versa.
Like the cell vector moduli [Fig.~(\ref{fig:cell-axis})], the volume
expands from the 0~K value to a slightly larger value appropriate to the
temperature of the simulation, around which its value oscillates. The
internal pressure, on the other hand, oscillates around 0~GPa, the value
fixed for the external pressure in this simulation. At the end of the
trajectory, the thermal averages of the internal pressure and the 
temperature of the system were computed to be 0.2~GPa and
999.7~K respectively, in good agreement with the imposed external values.

\subsection{Thermal expansion of silicon}
\label{sub:silicon}

\begin{figure}[t]
\begin{center}
\leavevmode
\epsfxsize=8.5cm
\epsffile{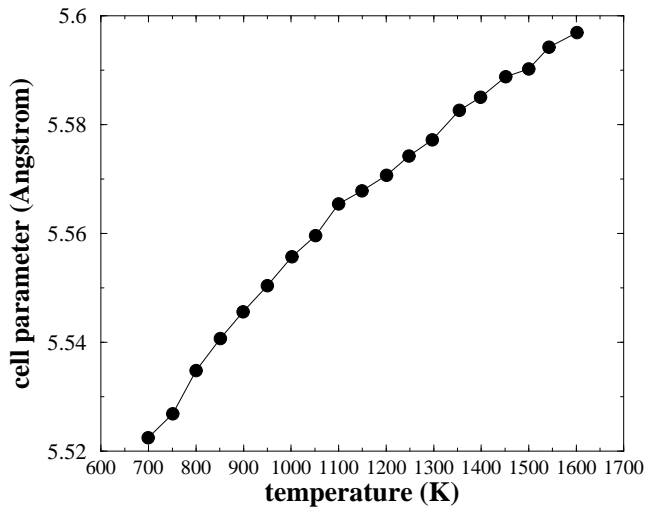}
\end{center}
\caption{Silicon cell parameter as a function of temperature obtained
from NPT MD simulations at 0~GPa external pressure and a series of
temperatures.}
\label{fig:cell_vs_T}
\end{figure}

Silicon has a smaller Debye temperature than diamond, c.a.~640~K, and I
will therefore use classical MD to study its thermal expansion
above this temperature. A series of NPT simulations at 0~GPa external
pressure and different temperatures have been carried out. The total length
of the simulation was 50~ps, using a time step of 1~fs. The external 
temperature of the simulation was varied in steps of 50~K between 700
and 1600~K, and at each such temperatures a simulation was conducted.
In all of them the value of $H_{\rm\scriptsize NPT}$ was accurately
conserved, and no appreciable drift in its value was observed. The variation
of the silicon cell parameter with temperature is illustrated in 
Fig.~(\ref{fig:cell_vs_T}). Qualitatively, two approximately linear 
behaviors can be observed, between 700 and 1150~K, and from
1150 to 1600~K, with the second range of temperatures having a slightly
lower slope. As can be seen, the behavior is not very smooth, indicating
that there is still a degree of statistical noise in the thermal averages
of the cell parameter computed from these simulations. I will dwell on the
possible causes of this below.

In principle, from the data shown in Fig.~(\ref{fig:cell_vs_T}) it would
be possible to calculate the thermal expansion coefficient, $\alpha$, of silicon
in the range of temperatures considered, but the statistical inaccuracies 
present in the data make this a difficult task. The experimental value
at 1000~K is $\alpha_{exp} = 4.3 \times 10^{-6} \mbox{K}^{-1}$, while from 
the data in the figure one can estimate a value of 
$\alpha_{calc} = 8.1 \times 10^{-6} \mbox{K}^{-1}$ at this temperature.
The thermal expansion coefficient is extremely sensitive, and given the
statistical uncertainties present in the results, the calculated value
can only be considered a rather crude estimate of the value predicted
by the TB model used; a more accurate estimate could be closer to the 
experimental value.

It is instructive to consider the possible causes of the poor statistics
observed in Fig.~(\ref{fig:cell_vs_T}). It is certainly not due to 
inadequate conservation of $H_{\rm\scriptsize NPT}$, which is 
sufficiently well conserved throughout all the simulations. A more likely
explanation is that the dynamics generated is not sampling the NPT 
ensemble with sufficient efficiency. Inefficient sampling can occur 
because some degrees of freedom do not easily exchange energy with 
the rest of the system, a situation which takes place when there are
largely different frequencies present. Fig.~(\ref{fig:volume}) lends
some weight to this consideration. There it can be seen that the volume
is oscillating quasi-harmonically with a single dominant frequency. The
internal pressure, however, has two dominant frequencies: a high one,
reflecting the thermal vibrations of the atoms, and a lower one, with
the same frequency as the volume. Sampling efficiency could be increased
by reducing the difference between the high frequency oscillations in 
the internal pressure and the frequency of the volume motion, which 
can be achieved using a lower barostat fictitious mass. It should be 
pointed out that, formally at least, the NPT ensemble is sampled 
independently of the values used for the masses of the fictitious degrees
of freedom, provided the dynamics is ergodic. The sampling efficiency,
however, does depend on the values chosen, and therefore this choice
must be made with care.

\subsection{Silicon under uniaxial external stress}
\label{sub:uniaxialstress}

In this final example, the capability of the method to cope with the 
simulation of systems subjected to non-hydrostatic external pressures
will be illustrated. Again, a 64~atom silicon supercell in the diamond
structure was used as test case. The external temperature was fixed at
1000~K, and a series of simulations were performed applying an 
external stress in the $[100]$ direction, varying from -5 to 5~GPa
in steps of 1~GPa. In the sign convention used here a positive sign
indicates a compressive pressure.

\begin{figure}[t]
\begin{center}
\leavevmode
\epsfxsize=8.5cm
\epsffile{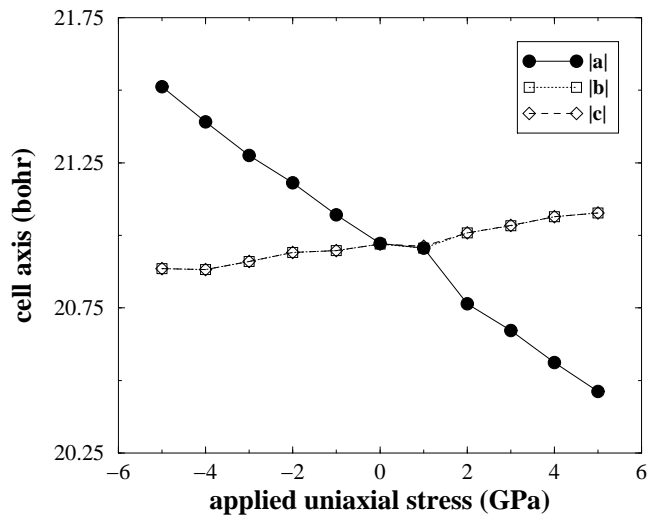}
\end{center}
\caption{Silicon cell vector lengths as a function of the applied uniaxial
external stress in the $[100]$ direction
from simulations at 1000~K external temperature.}
\label{fig:pressure}
\end{figure}

In Fig.~(\ref{fig:pressure}) the variation of the cell axis as a function
of the applied stress is shown. The supercell length in the direction of 
the applied stress has a negative slope, and it starts at an expanded
value, larger than the equilibrium value at the same temperature and
zero stress, reaching a value below this when the applied stress is
compressive. The other two cell lengths have positive, though smaller, slopes,
expanding as the first dimension is compressed. Both $|\mbox{\bf b}|$ and
$|\mbox{\bf c}|$ show nearly identical behavior, as expected, given the
symmetry of the system. This figure illustrates how 
the elastic constants of materials (in this case the Poisson ratio)
could be evaluated at finite temperatures using this methodology.

\section{Conclusions}
\label{sec:conclusions}

In this paper I have presented a stable, symplectic algorithm for integrating
the Hamiltonian equations of motion resulting from the combination of the 
Souza and Martins~\cite{souza:martins} metric tensor-based constant-pressure
scheme and the Nos\'{e}-Poincar\'{e} thermostat scheme of 
Bond and coworkers~\cite{bond:leimkuhler:laird}. The dynamics generated by
these equations samples the isothermal-isobaric (NPT) ensemble with full-cell
flexibility. Conditions of non-hydrostatic external pressure can also be
simulated. The numerical scheme advocated here is easy to implement in
existing molecular dynamics codes. The capabilities of this methodology 
have been illustrated with a series of numerical experiments in carbon
and silicon in the diamond structure, using a tight binding model.

\begin{acknowledgments}

It is a pleasure to acknowledge 
J.~L.~Mozos, P.~Ordej\'{o}n and J.~L.~Martins for helpful discussions.

\end{acknowledgments}

\end{document}